\documentclass[sigconf,nonacm,natbib=true]{acmart}
\AtBeginDocument{%
  }

\usepackage{pbalance}
\usepackage{booktabs}
\usepackage{multirow}
\usepackage{threeparttable}
\usepackage{tabularx}
\usepackage{url}

\newcolumntype{Y}{>{\centering\arraybackslash}X}

\makeatletter
\newcommand{\acmrightssize}{\fontsize{8}{9.5}\selectfont}
\newcommand{\firstpagerights}[1]{%
  \begingroup
    \renewcommand\thefootnote{}%
    \footnotetext{%
      \acmrightssize
      \raggedright
      \setlength{\parskip}{0pt}%
      \setlength{\parindent}{0pt}%
      #1%
    }%
    \addtocounter{footnote}{0}%
  \endgroup
}
\makeatother

\begin{document}

\title[Metadata, Structure, or Strategy?]{Metadata, Structure, or Strategy?\\ A Decomposition of RAG Context Enrichment}

\author{Saber Zerhoudi}
\orcid{0000-0003-2259-0462}
\affiliation{%
  \institution{University of Passau}
  \city{Passau}
  \country{Germany}
}
\email{saber.zerhoudi@uni-passau.de}

\author{Michael Granitzer}
\orcid{0000-0003-3566-5507}
\affiliation{%
  \institution{University of Passau}
  \city{Passau}
  \country{Germany}
}
\affiliation{%
  \institution{Interdisciplinary Transformation University Austria}
  \city{Linz}
  \country{Austria}
}
\email{michael.granitzer@uni-passau.de}

\author{Jelena Mitrovi\'{c}}
\orcid{0000-0003-3220-8749}
\affiliation{%
  \institution{University of Passau}
  \city{Passau}
  \country{Germany}
}
\email{jelena.mitrovic@uni-passau.de}

\renewcommand{\shortauthors}{Zerhoudi et al.}

\begin{abstract}
Retrieval-augmented generation (RAG) systems increasingly enrich retrieved passages by attaching quality \emph{metadata}, \emph{structuring} them into explicit records, and adopting multi-hop retrieval \emph{strategies} that accumulate evidence across steps. These changes assume that richer context yields better answers, yet existing evaluations cannot test this because they vary all three factors at once. We isolate each factor in a controlled experiment across six benchmarks, four models from three families, and five enrichment levels, totaling over 24,000 evaluated responses. The assumption does not hold. Most enrichment reduces accuracy. Models prompted to use confidence scores comply correctly yet produce worse answers, a gap between utilization and accuracy that no prior work has measured. What determines answer quality is not how much metadata the context carries but whether the model can act on it for the given task. When metadata and retrieval strategy are aligned with model capabilities, a smaller model outperforms a frontier model by 19~F1 points. These findings motivate a processability hierarchy that predicts, from pre-training properties alone, which metadata a model can productively use, reframing RAG design as a question of model-context alignment rather than metadata accumulation.
\end{abstract}

\keywords{Retrieval-augmented generation, Quality metadata, RAG evaluation, Retrieval strategies, Context enrichment}

\maketitle
\enlargethispage{2\baselineskip}
\firstpagerights{%
  \textcopyright{} Springer Nature Switzerland AG, 2026. This is the author's accepted manuscript of a paper accepted for publication at the European Conference on Machine Learning and Principles and Practice of Knowledge Discovery in Databases (ECML-PKDD 2026).\\
  The Version of Record will be published in the conference proceedings in Springer \emph{Lecture Notes in Computer Science (LNCS)}; the DOI will be added here once it is available.\\
  This author's version is made available in accordance with the Springer Nature Accepted Manuscript terms of use: \url{https://www.springernature.com/gp/open-research/policies/accepted-manuscript-terms}.
}

%----------------------------------------------------------------------
% 1. INTRODUCTION
%----------------------------------------------------------------------
\section{Introduction}

Retrieval-augmented generation (RAG) improves large language models by grounding them in retrieved external knowledge~\cite{Lewis:2020:NeurIPS}, effectively serving as an external memory that supplements what the model learned during training. A common assumption is that richer memory should produce better answers: systems therefore attach metadata such as timestamps, confidence scores, or provenance chains, reformat passages into structured records, and introduce more sophisticated retrieval procedures~\cite{Li:2025:ACMMM,Lin:2025:arXiv,Sanmartin:2024:arXiv}. But richer context is not inherently better. A model can correctly process a piece of metadata and still produce a worse answer, and a cleaner representation can change reasoning behavior even when no new information is added. This suggests that the central question is not simply whether context is richer, but which parts of retrieved context a model can actually use.

Existing RAG evaluations make this question difficult to answer because they typically change three factors at once: the \emph{metadata} attached to retrieved evidence (temporal validity, confidence, provenance), the \emph{structure} of the input (converting passages into JSON records or Markdown tables), and the \emph{strategy} of retrieval itself (decomposing questions into sub-queries or reusing successful search plans). Reported gains are measured against a raw-passage baseline, but no prior evaluation includes a structured-but-metadata-free control. It therefore remains unclear whether improvements come from the added metadata, from the representation format, or from the retrieval procedure.

This confound matters because different metadata types plausibly place different demands on the model. Temporal validity reduces to date comparison, an operation models encounter constantly in training data, while confidence scores or provenance may be visible without being productively usable for answer selection. A model can also be drawn toward easy-to-process metadata that is irrelevant to the task. Under this view, RAG quality depends not only on retrieval quality, but on \emph{model-context alignment}: whether the representation of retrieved evidence matches operations the model can perform reliably.

We study this question through a controlled decomposition of RAG context enrichment along three axes: metadata, structure, and strategy. We compare raw passages against a structured-but-metadata-free control, then incrementally add four metadata types and vary the retrieval procedure. The evaluation spans six benchmarks, four models from three families, and five enrichment levels, totaling over 24{,}000 evaluated responses. This design lets us isolate whether gains arise from metadata itself, from formatting effects, or from smarter retrieval.

Our contributions are threefold. First, we introduce a controlled evaluation that separates metadata, structure, and retrieval strategy, including a structured-but-metadata-free control absent from prior work. Second, we show that adding more metadata reduces accuracy on every benchmark: structure alone degrades reasoning, temporal metadata helps only date filtering, and confidence creates a utilization-accuracy gap in which correct processing leads to worse answers. Third, we propose a hypothesis-generating processability hierarchy that reframes RAG design as a problem of model-context alignment rather than metadata accumulation.

\noindent We release all schemas, evaluation code, and data.\footnote{\url{https://github.com/searchsim-org/ecml26-nuggetpedia-eval}; supplementary materials contain full specifications, additional tables, and reproducibility details.}

%----------------------------------------------------------------------
% 2. RELATED WORK
%----------------------------------------------------------------------
\section{Related Work}

\paragraph{Quality Metadata in RAG.}
Standard RAG retrieves passages by semantic similarity with no indication of reliability, currency, or sourcing~\cite{Lewis:2020:NeurIPS}. Recent systems address these limitations by attaching quality metadata, each tackling one dimension at a time.

T-GRAG~\cite{Li:2025:ACMMM} adds temporal conflict resolution via temporal knowledge graphs, outperforming GraphRAG~\cite{Edge:2025:arXiv} by 22--47 points on temporally sensitive queries. Freshness- and scope-aware web retrieval~\cite{Zerhoudi:2026:arXiv} similarly conditions results on recency for LLM assistants. EVOREASONER~\cite{Lin:2025:arXiv} adds confidence scoring and temporal evolution graphs, enabling an 8B model to match a 671B on temporal reasoning. KG-RAG~\cite{Sanmartin:2024:arXiv} organize knowledge as graphs for multi-hop reasoning. Closest to our setting, NuggetIndex~\cite{Zerhoudi:2026:SIGIR} stores knowledge as governed atomic nuggets for maintainable RAG; we adopt that atomic representation but, rather than proposing an indexing method, decompose the separate contributions of metadata, structure, and strategy. Each system reports gains over raw-passage baselines, but none isolates individual metadata contributions (e.g., how much does confidence add \emph{beyond} temporal validity?), and none includes a structured-but-metadata-free control to rule out formatting effects. Our design fills both gaps: it tests each metadata type's marginal value while including the missing G1 control that separates structure from metadata.

\paragraph{Retrieval Strategy.}
A complementary line of work controls the \emph{retrieval process} rather than the retrieved content. IRCoT~\cite{Trivedi:2023:ACL} interleaves retrieval with chain-of-thought steps. Self-RAG~\cite{Asai:2023:ICLR} trains the model to decide when to retrieve and whether retrieved content is relevant. Adaptive-RAG~\cite{Jeong:2024:NAACL} classifies query complexity and routes simple queries to non-retrieval pathways. Agent-based approaches such as PersonaRAG~\cite{Zerhoudi:2024:IRRAG} add user-centric agents that refine retrieved context before generation. Decomposing a query into a sequence of sub-queries is also related to modeling how users reformulate queries across a search session~\cite{Zerhoudi:2021:FIRE,Zerhoudi:2025:CIKM}. These systems decide \emph{when} and \emph{how} to fire retrieval; we decide \emph{what the model receives} when retrieval fires. Storing retrieval strategies as \emph{reusable knowledge} is less explored. Memento~\cite{Zhou:2025:arXiv} stores successful problem-solving traces and replays them, but its case bank is private to a single agent. MetaRAG~\cite{Zhou:2024:WWW} and AlignRAG~\cite{Wei:2025:arXiv} generate adaptive strategies during generation but discard them after each session. No prior work studies collectively shared, openly refined retrieval recipes with negative knowledge, a gap that SearchNuggets address (Section~\ref{sec:searchnugget}).

\paragraph{Index Quality versus Model Scale.}
Recent evidence challenges the assumption that larger models always produce better answers. OpenScholar~\cite{Asai:2024:arXiv} shows an 8B model outperforming GPT-4o with domain-specific retrieval; longer context degrades accuracy on several benchmarks~\cite{Du:2025:arXiv}; and the Lost-in-the-Middle effect~\cite{Liu:2024:TACL} shows models systematically underweight information in the center of long contexts. We draw on prior evaluation frameworks~\cite{Es:2024:EACL,SaadFalcon:2024:NAACL} for metric design while noting that none decomposes ``quality'' into testable metadata dimensions.

%----------------------------------------------------------------------
% 3. METHOD
%----------------------------------------------------------------------
\section{Method}\label{sec:method}

\subsection{NuggetPedia}

The experiments require a platform that provides metadata enrichment and retrieval strategy as independently testable components. We built NuggetPedia~\footnote{\url{https://nuggetpedia.com}}, building on the governed atomic-nugget representation of NuggetIndex~\cite{Zerhoudi:2026:SIGIR}, to serve this role. Implementation choices (JSON schema, nugget extraction quality, confidence calibration) may modulate effect sizes, so we test across four models from three families and validate each primary finding with external sources (Wikidata metadata, MuSiQue native passages, gold decompositions; Section~\ref{sec:deconfound}). The platform organizes knowledge in two complementary structures: metadata-enriched nuggets and SearchNuggets.

\paragraph{Metadata-enriched nuggets.}
A metadata-enriched nugget is an atomic fact carrying four quality metadata types: temporal validity (when the fact is current, as \texttt{valid\_from}/\texttt{valid\_until} timestamps), confidence (calibrated reliability, 0.2--1.0), conflict (documented source disagreements), and provenance (source URL, extraction date, verification count). The G1 control uses the same atomic JSON format with blank quality fields, isolating formatting from metadata content.

\paragraph{SearchNuggets.}\label{sec:searchnugget}
A SearchNugget is a reusable retrieval recipe: how to decompose a query type, which strategies work, and what \emph{not} to try (negative knowledge). Unlike Memento's~\cite{Zhou:2025:arXiv} private case bank, SearchNuggets are shared and accumulate collective retrieval experience.
The collection was seeded from GAIA benchmark~\cite{Mialon:2023:ICLR} questions and expanded with negative knowledge and cross-domain decomposition patterns. The result is 1{,}194 SearchNuggets spanning 12 domains, all used in the experiments reported below. Nuggets were extracted using Claude Sonnet 4.6; SearchNuggets were seeded using ChatGPT-5.4~\cite{Singh:2025:arXiv} thinking traces. Both underwent human relevance validation before inclusion.\footnote{Full details in the supplementary material.}

\subsection{Experimental Design}

\paragraph{Metadata-richness levels (5 conditions).}
Each level adds one dimension of quality metadata to the previous level, creating a cumulative ladder that allows marginal contribution analysis. \textbf{G0}: raw passages, the standard RAG baseline with unprocessed text chunks that retain their original overlapping discourse context (no atomization). G0 is therefore the natural-text reference against which the atomic conditions below are compared. \textbf{G1}: atomic JSON with blank quality fields, the critical formatting control that isolates structure from metadata. \textbf{G2}: G1 plus temporal validity windows (\texttt{valid\_from}, \texttt{valid\_until}). \textbf{G3}: G2 plus confidence scores and conflict metadata. \textbf{G4}: G3 plus full provenance chains (source URL, extraction date, verification count).

\paragraph{Retrieval strategies (3 conditions).}
\textbf{S0}: single-shot top-$k$ retrieval by semantic similarity, the standard baseline. \textbf{S1}: LLM self-decomposition in a ReAct-style loop~\cite{Yao:2022:ICLR}, where the model generates sub-questions, retrieves for each, and synthesizes. \textbf{S2}: SearchNugget-guided retrieval, where the model receives a domain-specific decomposition template with suggested tool choices and negative knowledge, then materializes the template into executable queries.

\paragraph{Models (4 models, 3 families).}
Primary pair: GPT-4.1-mini (cost-efficient) and GPT-4.1 (frontier), both from OpenAI. Cross-family validation: Llama-3.1-8B-Instant (Meta, served via Groq) and Qwen3-32B (Alibaba, served via Groq). Four models from three independent families ensure that findings are not artifacts of a single architecture or training regime. All models receive the same system prompt that names metadata fields but does not instruct how to use them; it never says ``reject expired facts'' or ``weigh claims by confidence''. Any metadata utilization is emergent model behavior, not prompted compliance.

\paragraph{Benchmarks (6 datasets).}
\textbf{TempLAMA}~(34,963 temporal probes; token F1): temporal \emph{filtering} --- can models exploit validity windows for time-sensitive factual questions? \textbf{TimeQA}~\cite{Chen:2021:NeurIPS} (6,150 time-sensitive questions over Wikipedia paragraphs; token F1): temporal \emph{comprehension} --- a second temporal benchmark testing whether gains transfer from date filtering to paragraph-level reasoning. \textbf{MuSiQue}~\cite{Trivedi:2022:TACL} (2,417 multi-hop questions requiring 2--4 hops; token F1): compositional reasoning chains. \textbf{HotpotQA}~\cite{Yang:2018:EMNLP} (7,405 multi-hop questions with distractors; token F1): parallel evidence comparison. \textbf{FEVER}~\cite{Thorne:2018:NAACL} (evidence-based claim verification; exact match): fact-checking via claim--evidence comparison. \textbf{SimpleQA}~\cite{Wei:2024:arXiv} (4,326 factual questions; exact match): negative control where metadata should have no effect. The six benchmarks span five task types, enabling us to test whether metadata value is universal or task-conditional. TempLAMA and MuSiQue receive the full five-level ladder; the remaining benchmarks are tested at anchor points (G0, G1, G4).

\paragraph{Stability and reproducibility.}
We use a hierarchical design: broad factorial screening ($n{=}50$ per cell, three seeds, 10{,}445 responses) followed by powered confirmation ($n{=}500$) on targeted experiments. Full significance tables and bootstrap CIs appear in the supplementary material.

%----------------------------------------------------------------------
% 4. RESULTS
%----------------------------------------------------------------------
\section{Results}\label{sec:results}

We report seed-42 results for the GPT-4.1 family unless otherwise noted; tables report pilot values with all primary effects confirmed at $n{=}500$ ($p{<}0.001$). Cross-model and cross-seed details in the supplementary material.

\subsection{Isolating Structure from Metadata}

\begin{table*}[t]
\caption{Marginal contribution of each metadata type ($\Delta$ over previous level, S0). On every benchmark, accuracy peaks before full enrichment; even TempLAMA's gain is driven entirely by one layer (temporal validity). TempLAMA and MuSiQue: GPT-4.1 family avg, 3-seed $\pm$ SE; others: GPT-4.1-mini.}\label{tab:marginal}
\centering\small
\begin{tabularx}{\textwidth}{@{}l*{6}{Y}@{}}
\toprule
Transition & TempLAMA & MuSiQue & HotpotQA & SimpleQA & FEVER & TimeQA \\
\midrule
G0 (base) & 0.717\,{\tiny$\pm$0.004} & 0.315\,{\tiny$\pm$0.005} & 0.489 & 0.660 & 0.708 & 0.188 \\
\midrule
G0$\to$G1 & $-$0.008\,{\tiny$\pm$0.006} & $-$0.078\,{\tiny$\pm$0.007} & $-$0.040 & $-$0.020 & \textbf{+0.012} & $-$0.035 \\
G1$\to$G2 & \textbf{+0.220}\,{\tiny$\pm$0.008} & $-$0.013\,{\tiny$\pm$0.008} & --- & --- & 0.000 & $-$0.010 \\
G2$\to$G3 & $-$0.005\,{\tiny$\pm$0.010} & +0.010\,{\tiny$\pm$0.010} & --- & --- & $-$0.014 & --- \\
G3$\to$G4 & $-$0.011\,{\tiny$\pm$0.008} & \textbf{+0.049}\,{\tiny$\pm$0.008} & --- & --- & $-$0.004 & --- \\
\midrule
G0$\to$G4 & +0.196 & $-$0.032 & $-$0.063 & $-$0.010 & $-$0.006 & $-$0.036 \\
\bottomrule
\end{tabularx}
\end{table*}

Table~\ref{tab:marginal} shows the marginal contribution of each metadata type across the four primary benchmarks. The G0$\to$G1 transition to structured-but-metadata-free JSON is negative on every benchmark except FEVER ($+$0.012), where structured fields aid claim--evidence comparison. Structured formatting alone hurts reasoning tasks. The penalty is format-dependent: replacing JSON with Markdown at G1 eliminates the structure penalty on MuSiQue ($+$0.086, $p{<}0.001$) while TempLAMA remains unchanged. This rules out the hypothesis that prior gains come from cleaner formatting: any metadata must overcome the formatting penalty to deliver net improvement.

With the confound controlled, the decomposition reveals a sharp task split. The entire G0$\to$G4 improvement on TempLAMA ($+$0.196) is driven by a single layer: temporal validity (G1$\to$G2: $+$0.220$\pm$0.008). Temporal windows account for 112\,\% of the total improvement, meaning subsequent layers (confidence, conflict, provenance) slightly offset the temporal gain. This makes sense: when a model knows a fact was valid from 2020 to 2023, it can directly answer ``Who did Brady play for in 2021?'' without inferring temporal context from prose.

A second temporal benchmark sharpens this finding. TimeQA~\cite{Chen:2021:NeurIPS} presents time-sensitive questions grounded in Wikipedia paragraphs --- the same domain as TempLAMA, with the same \texttt{valid\_from}/\texttt{valid\_until} metadata. Yet the effect reverses: G2 degrades F1 by $-$0.045 relative to G0. The difference is operational: TempLAMA requires date \emph{filtering}, a single-nugget operation that temporal windows directly enable; TimeQA requires \emph{comprehending} interleaved temporal facts across a passage, where atomization destroys the narrative continuity needed for reasoning. The processability hierarchy (Section~\ref{sec:hierarchy}) predicts this: date comparison is directly processable; paragraph comprehension is not. Including TimeQA and FEVER (claim verification, G0$\to$G4: $-$0.006), adding more metadata reduces accuracy on every benchmark: even TempLAMA peaks at G2, with subsequent layers offsetting the temporal gain.

On multi-hop QA, the picture inverts. The total G0$\to$G4 effect on MuSiQue is negative ($-$0.032), and HotpotQA confirms this pattern with a larger negative ($-$0.063). Provenance is the only positive contributor on MuSiQue ($+$0.049$\pm$0.008), though an ablation (Section~\ref{sec:provenance}) reveals that ${\sim}$40\,\% of this gain comes from removing a noisy \texttt{nugget\_id} field rather than from source metadata itself. The remaining ${\sim}$60\,\% is a genuine source-attribution effect, but it cannot overcome the cumulative drag of formatting ($-$0.078) and irrelevant temporal metadata ($-$0.013). SimpleQA shows no metadata effect ($-$0.010), as expected: factual recall questions either have the answer in context or they do not, and quality metadata cannot create information that is absent.

\paragraph{The optimal format is minimal.}
If temporal validity drives the gains, does it need JSON wrapping? Four serialization schemes (JSON, XML, natural language, key=value) all beat G0 on TempLAMA ($p{<}0.05$). Replacing JSON with Markdown serialization (\texttt{\#\#\# heading} + \texttt{- **key**: value}) goes further: G4-md matches or exceeds G4-JSON on both primary benchmarks ($+$0.013 on TempLAMA, $+$0.037 on MuSiQue, $p{=}0.001$) while using 33--46\,\% fewer input tokens. On MuSiQue, Markdown rescues 37 questions where JSON scored F1\,=\,0 (vs.\ 13 reverse) and halves the formatting penalty relative to G0 ($-$0.040 vs.\ $-$0.077). Temporal information drives the benefit; delivery format has a smaller but consistent effect, with lighter formats outperforming heavier ones.

\subsection{Isolating Strategy from Metadata}

Table~\ref{tab:retrieval} shows that metadata and strategy address different tasks. On MuSiQue, S2 provides $+$0.090 over S0 at G0, a larger effect than any metadata transition. The S2 benefit is positive in all 10 model$\times$enrichment-level cells on MuSiQue across the pilot ($n{=}50$, 3 seeds $\times$ 2 models $\times$ 3 levels).

\begin{table}[t]
\caption{Retrieval strategy effectiveness (averaged across GPT-4.1 models, $n{=}50$).
S0 and S2 are 3-seed means; S1 is seed~42 (validation run). S2 (SearchNugget-guided) dominates multi-hop; S1 (self-decomposition) performs best on temporal probes.}
\label{tab:retrieval}
\centering\small
\begin{threeparttable}
\begin{tabularx}{\columnwidth}{@{}l*{4}{Y}@{}}
\toprule
 & \multicolumn{2}{c}{TempLAMA (F1)} & \multicolumn{2}{c}{MuSiQue (F1)} \\
\cmidrule(lr){2-3}\cmidrule(lr){4-5}
Strategy & G0 & G4 & G0 & G4 \\
\midrule
S0 --- Standard top-$k$        & 0.717 & 0.913 & 0.315 & 0.282 \\
S1 --- Self-decomposition$^\dagger$ & 0.769 & 0.914 & 0.320 & 0.284 \\
S2 --- SearchNugget            & 0.748 & 0.903 & \textbf{0.405} & \textbf{0.333} \\
\bottomrule
\end{tabularx}
\begin{tablenotes}[flushleft]
\footnotesize
\item$^\dagger$ Single seed (validation run); all others are 3-seed means.
\end{tablenotes}
\end{threeparttable}
\end{table}

\paragraph{Scale-up and capability interaction.}
At $n{=}500$, the strategy effect strengthens and differentiates by model capability: GPT-4.1 gains more from S2 than GPT-4.1-mini ($+$0.129 vs.\ $+$0.053 at G4, both $p{<}0.001$). This inverts the temporal finding: weaker models benefit more from informational metadata (temporal validity), while stronger models benefit more from retrieval templates. The combined effect is substantial: GPT-4.1 with G4+S2 achieves 0.430, a net $+$0.095 over the G0/S0 baseline ($p{<}0.001$).

On TempLAMA, the pattern reverses. S1 (self-decomposition) slightly outperforms S2 at both G0 and G4. This makes sense: temporal probes are single-hop questions (``Who was the president in 2019?'') that do not benefit from multi-step decomposition templates. The LLM's ad-hoc reasoning in S1 is sufficient; rigid retrieval templates add overhead without value.

Raw SearchNugget templates used directly as queries perform worse than S1; adding LLM materialization, where the model converts structural hints into effective queries, accounts for 92\% of the S2 improvement.

\paragraph{Template quality, not template source.}
Does the benefit depend on NuggetPedia's specific templates? Replacing SearchNuggets with MuSiQue's gold decompositions yields $+$0.152 over S0 ($p{<}0.001$), larger than any SearchNugget result. LLM self-decomposition (S1) produces no improvement. Any decomposition does not help; only good decomposition does. Template-guided retrieval is a general mechanism, not a platform-specific artifact.

Metadata and strategy are not merely independent: they interact. Table~\ref{tab:retrieval} shows that S2's advantage over S0 on MuSiQue shrinks from $+$0.090 at G0 to $+$0.051 at G4: metadata partially addresses the same retrieval failures that SearchNuggets solve, reducing their marginal value. Conversely, S2 at G0 (0.405) outperforms S0 at G4 (0.282) by 12 points, suggesting that strategy can fully compensate for missing metadata on multi-hop tasks, while metadata alone cannot compensate for poor retrieval.

The S2 benefit is task-dependent. On HotpotQA ($n{=}500$), S2 hurts GPT-4.1-mini ($-$0.070 at G0, $p{<}0.001$). The divergence reflects structural differences: MuSiQue requires sequential compositional chains that align with decomposition templates, while HotpotQA requires parallel evidence comparison. The strategy claim holds for compositional multi-hop tasks, not multi-hop tasks generally.

With both factors isolated, the final piece of the decomposition asks whether combining them can close the gap between a small model and a large one.

\subsection{When Metadata and Strategy Align, Does Scale Still Matter?}

External enrichment may substitute for model capability. If metadata and retrieval templates supply what a weaker model cannot compute on its own, then weaker models should benefit more from enrichment than stronger ones. We test this across four models from three independent families to ensure the effect is not an artifact of a single architecture.

\begin{figure}[t]
\centering
\includegraphics[width=\columnwidth]{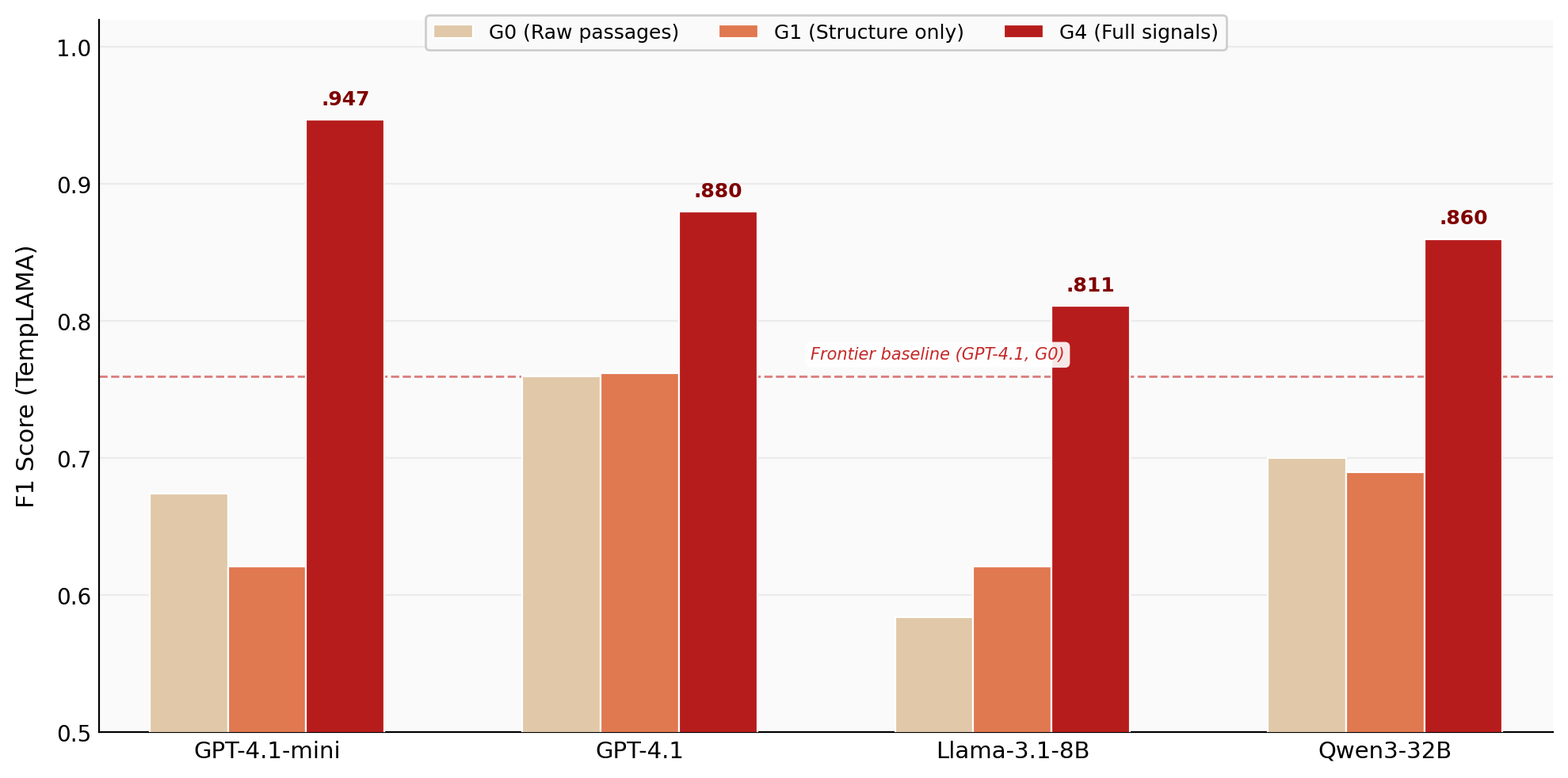}
\caption{Cross-family metadata effect on TempLAMA (S0). Each cluster shows three governance levels: raw passages (G0), structure only (G1), full signals (G4). GPT-4.1: 3-seed means; Llama and Qwen: seed~42. Dashed line: frontier model's raw-passage baseline. With full metadata, every model surpasses it; weaker models gain most.}\label{fig:hero}
\end{figure}

Figure~\ref{fig:hero} confirms the hypothesis across three model families and four models spanning a wide capability range. Adding metadata (G0$\to$G4) improves all four models on TempLAMA, and the effect is inversely related to model capability: GPT-4.1-mini gains the most ($+$0.273), followed by Llama-3.1-8B ($+$0.227) and Qwen3-32B ($+$0.160), with the frontier model GPT-4.1 gaining the least ($+$0.120). The models with fewer internal resources benefit most from external metadata, consistent with enrichment substituting for model capability.

GPT-4.1-mini with full metadata (0.947, 3-seed mean) outperforms GPT-4.1 with raw passages (0.760) by nearly 19~points. A small model with the right index beats a frontier model without it, at a fraction of the per-query cost. Even the 8B Llama model with full metadata (0.811) exceeds GPT-4.1-mini without them (0.674), demonstrating that the effect holds across family boundaries.

The G1 bars confirm the structure finding across families: three of four models show near-zero or negative G0$\to$G1 deltas, and no model shows structural gains large enough to explain the G0$\to$G4 improvements.

On MuSiQue, metadata alone produces near-zero change across all families, and the direction varies by architecture (Qwen: $-$0.060; GPT-4.1: $+$0.014). Combining metadata with SearchNugget-guided retrieval (S2 at G4) yields the best multi-hop performance for GPT-4.1, but open-weight models show the opposite: Llama and Qwen produce lower-quality search templates, and S2 hurts their performance. Strategy value depends on the materializing model's instruction-following quality.

The results so far show \emph{what} happens when metadata, structure, and strategy are varied. They do not explain \emph{why} structure hurts, why temporal metadata dominates, or why models can process metadata correctly and still get a worse answer. The next section traces all three to the same cause.

%----------------------------------------------------------------------
% 5. ANALYSIS
%----------------------------------------------------------------------
\section{Analysis}\label{sec:analysis}

\subsection{Why Structure Degrades Accuracy}\label{sec:g1}

Structure alone degrades performance across benchmarks and all four model families. Understanding why helps explain when metadata must work hardest to deliver net gains.

Atomizing passages into JSON removes discourse connectives and narrative flow that models exploit for chain reasoning. For temporal probes, the loss is minimal ($-$0.008) because the temporal validity field compensates. For multi-hop questions, it is substantial: MuSiQue ($-$0.078) and HotpotQA ($-$0.040) require bridging inferences between facts that atomization destroys. The cleanest test is G1: replacing JSON with Markdown removes 28--36\,\% of input tokens and eliminates the MuSiQue structure penalty ($+$0.086, $p{<}0.001$), while TempLAMA is unaffected. At G4, Markdown's advantage persists (33--46\,\% fewer tokens, $+$0.037 on MuSiQue), confirming the penalty compounds with metadata syntax.

A concrete case illustrates: for ``As of 2012, Peyton Manning plays for \_X\_?'', the G0 context contains contradictory passages. GPT-4.1-mini picks the first (Colts, F1\,=\,0.00), a position-bias failure; at G2, the temporal window resolves it (F1\,=\,1.00). The frontier model answered from parametric knowledge, but the cost-efficient model needed the metadata.\footnote{Structure-induced refusal example in the supplementary material.} Smaller models are more susceptible to both position bias (helped by metadata) and structure deference (hurt by it).

\subsection{How Models Process Metadata}\label{sec:behavioral}

The results above show \emph{what} happens: temporal metadata helps, confidence does not, structure hurts. This section examines \emph{why} by classifying how models actually engage with metadata in their chain-of-thought. We classified 2{,}400 G4 chain-of-thought responses across three model families (GPT-4.1-mini: 1{,}200; Llama-3.1-8B: 800; Qwen3-32B: 400) for explicit metadata utilization using an LLM classifier validated at $\kappa{=}0.97$ against 69 human annotations.\footnote{Per-metadata agreement ($\kappa{=}0.92$--$1.00$) and classifier validation in the supplementary material.}

\begin{table}[t]
\caption{How often models cite each metadata signal in their chain-of-thought, naturally vs.\ when prompted (1{,}200 G4 responses, GPT-4.1-mini, $\kappa{=}0.97$). Prompting boosts confidence citations from 14\,\% to 70\,\% but \emph{reduces} accuracy.}
\label{tab:behavioral}
\centering\small
\begin{threeparttable}
\begin{tabularx}{\columnwidth}{@{}lYY@{}}
\toprule
Metadata Signal & Natural & Prompted \\
\midrule
Temporal   & 50\% & 50\% \\
Confidence & 14\% & 70\% \\
Conflict   & 15\% & 18\% \\
Provenance & 0\%  & 3\% \\
\bottomrule
\end{tabularx}
\begin{tablenotes}[flushleft]
\footnotesize
\item \textit{Accuracy $\Delta$:} TempLAMA $-$0.192 \quad MuSiQue $-$0.032.
\end{tablenotes}
\end{threeparttable}
\end{table}

\paragraph{Content-before-metadata attention boundary.}
Table~\ref{tab:behavioral} reveals a sharp boundary between content-level and meta-level metadata. Across all three model families, temporal utilization is consistently high (50--55\,\%) while provenance is effectively zero (0--6\,\%). Confidence and conflict utilization vary by architecture (confidence: 14--49\,\%; conflict: 0--15\,\%), suggesting these mid-level metadata types interact with model-specific training. The architecture-independent finding is the boundary itself: models spontaneously attend to metadata that constrains the answer but are blind to attribution metadata, even when source names are explicitly present in context. The same pattern replicates across Llama-3.1-8B (800 responses) and Qwen3-32B (400 responses).

\paragraph{The utilization--accuracy gap.}
The ``Prompted'' column in Table~\ref{tab:behavioral} shows what happens when models are instructed to use metadata. Confidence references jump from 14\,\% to 70\,\% ($+$56pp) while conflict and provenance barely move ($+$3pp). The provenance barrier holds: even under explicit instruction to ``cite provenance for each claim'', utilization reaches only 3\,\% with 0\,\% appropriateness; the model \emph{cannot} extract actionable information from source names.

The accuracy impact is negative on both benchmarks: $-$0.192 on TempLAMA and $-$0.032 on MuSiQue. The mechanism is direct: the instructed prompt tells the model to weight claims by confidence, causing it to prefer high-confidence nuggets even when they are temporally wrong. In a concrete case, the model selects ``Peyton Manning plays for Indianapolis Colts'' (confidence\,=\,0.9, valid from 2010) over ``Denver Broncos'' (lower confidence, valid from 2012) for a question asking ``as of 2012,'' where confidence overrides temporal reasoning. Utilization is not accuracy: a model can process metadata correctly (correctness under prompting: confidence 94\,\%, temporal 88\,\%) and still produce a worse answer when the prompted metadata conflicts with the task-relevant one.

No prior work measures this gap between metadata utilization and downstream accuracy. We develop a candidate account below, which we present as a hypothesis rather than a validated mechanism.

\paragraph{Why correct processing hurts: metadata competition.}
This result is not a prompt engineering failure. We propose that it reflects how transformers allocate attention over context. We cannot confirm this account here, since we do not measure attention directly; we set out the prediction it makes and flag attention-level validation as future work. The account runs as follows. Attention across retrieved passages is finite. Directing it toward confidence metadata leaves less for temporal windows. Confidence scores use a familiar $[0,1]$ format that is easy for the model to engage with; temporal windows require comparing dates against the query, a harder operation. When both are present and a prompt activates confidence, the easier metadata captures more attention regardless of which one is more useful for the task. This extends the Lost-in-the-Middle effect~\cite{Liu:2024:TACL} from position to content type: models preferentially process information that narrows the answer space, regardless of position. Full enrichment (G4) degrades multi-hop QA by 5--8\,\% because it adds easy-to-process metadata that is not relevant to reasoning tasks. The optimal configuration (G0.5: plain-text temporal only) works precisely because it eliminates competition. There is only one metadata type, and it is the task-relevant one.

\subsection{Design Principles and Transferable Artifacts}

The metadata competition account and processability hierarchy jointly yield three principles for metadata-enriched RAG. First, \emph{less is more}: the optimal metadata set is the minimal set of task-relevant metadata, not the maximal set of available types, and even the serialization format should be as lightweight as possible (Markdown's Pareto dominance over JSON illustrates this at the token level). Second, \emph{which metadata to include matters more than metadata quality}: the decision should be based on task type, not on metadata fidelity. Well-calibrated confidence scores hurt temporal tasks just as much as poorly calibrated ones (our perturbation experiment confirms this: inverting all confidence scores has zero effect at $n{=}500$). Third, \emph{the easier-to-process metadata dominates}: when two metadata types coexist, the one the model can more readily engage with captures more attention (Section~\ref{sec:hierarchy} formalizes why). System designers should either remove the easier but less relevant metadata or develop attention-steering mechanisms. The practical reframing is from ``which metadata to add'' to ``which metadata to withhold''.

\subsection{A Processability Hierarchy for RAG Metadata}\label{sec:hierarchy}

The behavioral analysis shows that models engage with some metadata spontaneously and ignore others completely, regardless of prompting. This section proposes a hypothesis-generating framework for predicting which metadata will fall where, based on what language model training can and cannot learn.

\paragraph{Three tiers.}
We observe three levels of metadata processability, each with a distinct behavioral signature.

\emph{Directly processable} metadata is used spontaneously without prompting. Temporal validity (50\,\% unprompted utilization, stable when prompted) falls in this tier. The model already knows how to compare dates: news articles, encyclopedic entries, and timestamped web pages all require date comparison for accurate next-token prediction, so temporal reasoning is acquired as a byproduct. When a nugget says ``valid from 2012 to 2023'', the model can immediately check whether the query date falls in that range.

\emph{Latently processable} metadata is recognized but not applied until prompted. Confidence scores (14\,\% unprompted, 70\,\% prompted, 94\,\% appropriate) fall here. The $[0,1]$ format is familiar from training data, but acting on scores (``prefer 0.9 over 0.6'') was never a prediction target. Prompting supplies the missing instruction ($+$56pp).

\emph{Unprocessable} metadata is ignored even when prompted. Provenance (0\,\% unprompted, 3\,\% prompted, 0\,\% appropriate) falls here. Source names appear throughout training data, but evaluating source reliability was never a prediction target --- the model cannot map ``ESPN'' to a trust judgment.

The framework rests on three empirical observations (temporal, confidence, provenance), sufficient to motivate the taxonomy but not to validate it. Its value lies in generating testable predictions rather than confirming a theory.

\paragraph{A two-dimensional account.}
The three tiers follow from two properties of the training objective (Table~\ref{tab:hierarchy}): how \emph{frequent} is the underlying operation in pre-training data, and whether it is \emph{necessary} for accurate next-token prediction.

\begin{table}[t]
\caption{The processability hierarchy as a two-dimensional framework. Each quadrant predicts a distinct behavioral signature based on whether the underlying operation is frequent in pre-training and necessary for accurate next-token prediction.}\label{tab:hierarchy}
\centering\small
\begin{tabularx}{\columnwidth}{@{}lYY@{}}
\toprule
 & Prediction-necessary & Prediction-unnecessary \\
\midrule
Frequent operation & \textbf{Directly processable} & \textbf{Latently processable} \\
 & {\scriptsize (temporal validity, recency)} & {\scriptsize (confidence scores, ratings)} \\
\addlinespace
Infrequent operation & \textbf{Poorly processable} & \textbf{Unprocessable} \\
 & {\scriptsize (specialized unit conversion?)} & {\scriptsize (provenance, peer review status)} \\
\bottomrule
\end{tabularx}
\end{table}

Temporal comparison is both frequent and prediction-necessary (continuing ``In 2019, the president was\_\_'' requires date reasoning). Numeric comparison is frequent but not prediction-necessary. Source reliability evaluation is neither. The framework predicts a fourth category (prediction-necessary but rare): metadata that models process unreliably, such as specialized unit conversion.

\paragraph{Predictions and diagnostics.}
The framework generates falsifiable predictions:\footnote{Additional predictions and an information-theoretic reading in the supplementary material.} \emph{geographic proximity} should be directly processable (spatial reasoning is frequent and prediction-necessary); \emph{peer review status} should be unprocessable (mapping venue names to quality requires missing training signal). The framework also offers a practical diagnostic: before adding a new metadata type, locate it in Table~\ref{tab:hierarchy}. Top-left metadata can be added without prompt changes; top-right needs prompt engineering and risks competition with task-relevant metadata; bottom-right will not help without fine-tuning.

\subsection{Validation}\label{sec:deconfound}

\paragraph{The provenance paradox.}\label{sec:provenance}
The G3$\to$G4 transition improves MuSiQue by $+$0.049 despite 0\,\% explicit provenance utilization. An ablation decomposes the effect: ${\sim}$40\,\% comes from removing a noisy \texttt{nugget\_id} field and ${\sim}$60\,\% from adding source metadata. One hypothesis is that credible source names lower the model's abstention threshold, acting as implicit permission to draw on parametric knowledge when context is insufficient. On TempLAMA, where answers are present, provenance is pure noise (G3$\to$G4: $-$0.040).

\paragraph{Cross-source validation.}
Three experiments test whether the primary findings depend on NuggetPedia's specific implementation. \emph{Confidence perturbation}: inverting all confidence scores ($n{=}500$, 3{,}000 questions) has zero effect on accuracy, confirming that models do not reason about score values. \emph{External temporal metadata}: replacing NuggetPedia annotations with Wikidata validity periods produces the same direction of effect ($+$0.080 vs.\ $+$0.235, $p{=}0.0005$); the smaller magnitude reflects coarser date granularity. \emph{External passages}: applying JSON wrapping to MuSiQue's native paragraphs replicates the G1 penalty ($-$0.014). Each finding replicates with at least one external source, reducing the risk that the results are platform-specific artifacts.

%----------------------------------------------------------------------
% 6. CONCLUSION
%----------------------------------------------------------------------
\section{Conclusion}

The field assumes richer metadata improves RAG accuracy. Our results suggest the bottleneck is attention rather than information. The findings are consistent with a single account: models tend to process whatever is easiest to compute on, which is not always what the task needs. We frame this as a hypothesis that attention-level analysis should test directly.

Temporal metadata helps because date comparison is in the model's repertoire. Confidence scores hurt because processing them displaces the temporal reasoning that matters. Structure alone is a penalty, not a shortcut. The two interventions are complementary: temporal metadata aids weaker models on time-sensitive tasks, template-guided retrieval aids stronger models on compositional ones. Together, they let a cost-efficient model beat a frontier model by 19 points. The processability hierarchy (Table~\ref{tab:hierarchy}) predicts these outcomes and offers a diagnostic for deciding which metadata belongs in an index.

This points to a tension in RAG research. The field is building richer indexes, longer contexts, and denser metadata per passage. On every benchmark, accuracy peaks before full enrichment, even on TempLAMA, where subsequent layers offset the temporal gain, and on TimeQA, where temporal metadata itself hurts. A model that receives five metadata types and uses one is not under-informed; it is over-served. If the constraint is not what the model receives but what it can absorb, the question is not what to add to the index, but what to leave out.

\subsection{Limitations}

Several boundary conditions qualify these findings. The 1{,}194 SearchNuggets come from controlled curation, not a live multi-agent ecosystem; deployment would introduce gaming, drift, and variable strategy quality that may shift the 92\,\% materialization finding. SearchNugget value also depends on the materializing model: S2 reliably helps the GPT-4.1 family but hurts open-weight models ($-$6 to $-$36 points on multi-hop), likely because smaller models produce lower-quality search templates. The gold-template ceiling analysis ($+$0.152) confirms that the mechanism is template-guided retrieval broadly: any source of high-quality decomposition templates would produce similar effects. SearchNuggets' advantage over gold decompositions is that they encode negative knowledge and cross-domain patterns that dataset-specific templates lack, enabling error avoidance that gold sub-questions do not anticipate.

The experimental scope is bounded in three ways. Six benchmarks cover five task types; other domains may exhibit different profiles. Four instruction-tuned models from three families confirm architecture independence, but base models and reasoning variants (e.g., o4-mini) may differ. We also test only the smaller open-weight checkpoints (Llama-3.1-8B, Qwen3-32B); whether the limited-attention pattern persists in larger open-weight versions, or partly reflects parameter count, remains open. Scale-up experiments confirm all primary effects at $p{<}0.001$, but step-wise transitions (G2$\to$G3, G3$\to$G4) remain at pilot scale ($n{=}500$).

Three design choices merit discussion. First, the metadata-richness ladder is cumulative: each level adds metadata to the previous one, so the design measures the marginal value of adding confidence \emph{on top of} temporal, not confidence \emph{instead of} temporal. The confidence perturbation experiment ($n{=}500$, 3{,}000 questions) provides indirect evidence: inverting all confidence scores has zero effect on accuracy, suggesting confidence contributes nothing regardless of what other metadata is present. Second, utilization depends heavily on prompt design ($+$56pp for confidence under explicit instruction), a limitation affecting all metadata-enriched RAG research. S1 (LLM self-decomposition) uses a single seed as a validation baseline; the gold-template analysis confirms that template quality, not decomposition per se, drives the strategy effect.

Third, the enriched conditions (G1--G4) deliver evidence as atomic records, which removes the overlapping passage context that ordinary text RAG preserves. G0 retains that context, so the structure penalty we report is measured against natural overlapping text rather than against a strawman baseline. We do not, however, test an intermediate format that keeps overlapping passages while adding metadata. Our claim that structure alone degrades reasoning therefore holds for atomic delivery; whether a non-atomic enrichment format that preserves discourse continuity would avoid the penalty is an open question, and the external-passage replication (JSON wrapping of MuSiQue's native paragraphs, $-$0.014) is only a first step toward answering it.

\subsection{Future Work}

Three directions follow naturally. \textbf{Task-adaptive metadata routing:} a preliminary experiment confirms the metadata competition account's prescriptive power: a few-shot router ($n{=}10$ examples) selecting governance level per query achieves $+$2.0pp over fixed G4, while a perfect per-question oracle gains $+$6.5pp, showing that per-query metadata \emph{withholding} outperforms fixed enrichment. Scaling with richer features and attention-level routing~\cite{Jeong:2024:NAACL} is a natural extension. \textbf{Attention-level validation:} the processability hierarchy predicts that temporal tokens receive disproportionate attention weight while provenance tokens are attended to like structural filler; direct measurement via attention attribution methods would validate or refine the framework. \textbf{Reasoning-aware atomization:} the negative G1 effect on multi-hop QA suggests that future formats should preserve inter-fact connections, perhaps through hyperedges or discourse markers that bridge atomic records without sacrificing metadata granularity.

\bibliographystyle{ACM-Reference-Format}
\bibliography{references}

\end{document}